\documentclass[doublecol,A4paper]{epl2}

\usepackage{graphicx}
\usepackage{amssymb}
\usepackage{booktabs}
\usepackage{multirow}
\usepackage{caption}

\title{Pair-pair interactions as a mechanism for high-T$_c$ \\ superconductivity}
\shorttitle{Pair-pair interactions}

\author{William Sacks\inst{1} \and Alain Mauger\inst{1} \and Yves Noat\inst{2}}
\shortauthor{W. Sacks, A. Mauger \& Y. Noat}

\institute{
\inst{1} Institut de Min\'{e}ralogie, de Physique des
Mat\'{e}riaux, et de Cosmochimie (IMPMC), UMR 7590,

\inst{2} Institut des Nanosciences de Paris (INSP), UMR 7588,\\

Sorbonne Universit\'{e}s, UPMC Paris 6, \\
4 place Jussieu, 75252 Paris Cedex 05, France }

\pacs{74.72.h}{First pacs description} \pacs{74.20.Mn}{Second pacs
description} \pacs{74.20.Fg}{Third pacs description}

\abstract{The mutual interaction between Cooper pairs is proposed as
a mechanism for the superconducting state. Above $T_c$, pre-existing
but fluctuating Cooper pairs give rise to the unconventional {\it
pseudogap} (PG) state, well-characterized by experiment. At the
critical temperature, the pair-pair interaction induces a Bose-like
condensation of these preformed pairs leading to the superconducting
(SC) state. Below $T_c$, both the condensation energy and the
pair-pair interaction $\beta$ are proportional to the condensate
density $N_{oc}(T)$, whereas the usual Fermi-level spectral gap
$\Delta_p$ is independent of temperature. The new order parameter
$\beta(T)$, can be followed as a function of temperature, carrier
concentration and disorder - i.e. the phase diagrams. The complexity
of the cuprates, revealed by the large number of parameters, is a
consequence of the {\it coupling of quasiparticles to Cooper-pair
excitations}. The latter interpretation is strongly supported by the
observed quasiparticle spectral function.}
\begin{document}

\maketitle

\section{I.\,Introduction}
\label{intro}

As Occam's razor would suggest, second  order phase transitions
often depend on few parameters\,\cite{stanley}. Such is the case for
the familiar magnetic, spin glass, charge-density wave, structural
transitions, etc., and the more exotic Kosterlitz-Thouless case for
two-dimensional systems\cite{nazarenko}. The superconducting phase
of `classical' materials follows this trend wherein a weak
attractive electron-electron interaction is responsible for the
transition. Essentially a single energy scale, the SC gap parameter
$\Delta_0$ at zero temperature, is relevant \cite{bcs}. In the
conventional theory of Bardeen, Cooper, Schrieffer (BCS) \cite{bcs}
pair-breaking quasiparticle excitations restore the normal-metal
state at the transition, such that $\Delta_0 = 1.76\,k_B\, T_c$.
Moreover, $\Delta_0$ fixes the scale of the critical currents, for
example in a Josephson junction, the upper critical field and the
coherence length, such as the vortex core radius.

In highly-correlated electron systems such as cuprates, many
physical properties of their SC state are not yet fully understood.
In addition to many theoretical ideas
\cite{rvb,rvbnum,preformedpairs,yeh,andrenacci,kivelson} the
experiments reveal an inherent complexity\,: a complex quasiparticle
self-energy\,\cite{normanKE,chubukov,eschrig}, coupling to a spin
collective
mode\,\cite{Arpes,campuzano,norman_ding,jenkins,castro,zazaSIS},
anomalous specific heat \cite{pgreview}, multiple energy scales
\cite{tallon,letacon,huscroft,sacks}, spatial inhomogeneities
\cite{pan,crendisorder,crenmaps,howald,mcelroy,sugimoto},
checkerboard oscillations \cite{hoffmann1,hoffmann2}, pseudogap
phenomena above
$T_c$\,\cite{pgalloul,ding,pgreview,yeh,millis,yazdani}, etc. The SC
state of cuprates is indeed far from the desired {\it lex
parsimoniae}.

The clear jump in magnitude of the physical parameters is also
striking. In addition to $T_c$, the very small coherence length,
large penetration depth, large Fermi-level gap value, small lower
critical-field, etc., come to mind. In the case of cuprates, the
precise link of the measured Fermi-level gap to the condensate is
still controversial. Consider BiSrCaCuO (2212) as a typical example.
The energy gap is $\Delta \simeq 32$ meV near optimal doping ($p$)
which, if directly tied to the critical temperature using the BCS
ratio, would give the erroneous value of $T_c \sim 220$ K. Moreover,
on the underdoped side of the phase diagram the energy gap increases
while $T_c$ decreases -- a clear paradox. Thus, $\Delta$ is not the
scale of $k_B\,T_c$.

The quasiparticle (QP) spectral function, expressing the material's
microscopic interactions, also contradicts the conventional picture
\cite{normanKE,chubukov,eschrig,zazaSIS,matsuda}. Measured by
tunneling or ARPES as a function of rising temperature, the cuprate
SC gap does not close at $T_c$ as would be expected, but rather a
{\it pseudogap} remains at the Fermi level. To within thermal
broadening, the pseudogap has about the same value as the SC gap,
with no coherence peaks, and it finally vanishes at the higher
temperature $\sim T^*$ \cite{rennerT}. An analogous PG exists in the
vortex core \cite{rennerB,phd,fischer_rev} or due to disorder
\cite{crendisorder,mcelroy}, where SC coherence is also lost. Thus
in the case of cuprates, the direct sign of the condensate is not
the gap {\it per se}, but instead the finer structure of the QP
spectral function (peak-dip-hump).

Very different interpretations have been proposed\,: the two-energy
scale gap function, implying a non-retarded pairing interaction
\cite{constraints,sacks,smn}, and the coupling to a spin collective
mode \cite{Arpes,chubukov,norman_ding,campuzano,normanPDH,
eschrig,jenkins,chi,zazaSIS,zaza3,ahmadi,castro}. Although
phenomenological, the static model describes very well the
particular signatures of the spectral function, the SC density of
states (DOS), and the transition to the pseudogap state
\cite{sacks}. Still, the origin of these signatures remains to be
clarified.

Many studies with scanning tunneling spectroscopy (STS)
\cite{pan,crendisorder,crenmaps,howald,mcelroy,sugimoto,fischer_rev},
showed real-space inhomogeneities even for pristine samples at low
temperature. According to conventional wisdom, without an applied
magnetic field or current, the SC order parameter should be
homogeneous. The phenomenon has led to in-depth theoretical studies
of the mechanisms underlying long-range SC order with particular
emphasis on the effects of disorder and percolation
\cite{ghosal,atkinson1,atkinson3,hirschfeld}. By studying a variety
of spectral data, we find that they have a distinctive shape on the
mesoscopic scale; the underlying parameters then become meaningful.

\vskip 1 mm

To summarize, a solution of the high-$T_c$ problem must account
for\,:
\parindent 0 pt

($i$) multiple energy scales and large spectral gap $\Delta_p$,

($ii$) large $T_c$, but paradoxically\,: $k_B\, T_c \ll \Delta_p$,

($iii$) the dome-shape of the $T-p$ phase diagram,

($iv$) the SC to pseudogap transition,

($v$) the unconventional QP dispersion (peak-dip-hump).

\parindent 12 pt

\vskip 1mm

In this work a microscopic theory is presented that addresses these
requirements. We assume the existence of pre-existing fluctuating
pairs above $T_c$ in the incoherent pseudogap
state\,\cite{kivelson,andrenacci,valdez,millis,huscroft,preformedpairs,yeh}.
Since the carrier-density and coherence length are small, a system
of interacting bosons is relevant\,; a new pair-pair interaction,
$\beta_{i,j}$, leads to a Bose-like condensation to the coherent SC
state at the critical temperature (section III).

Simple expressions for the gap equation and the condensation energy
are given and the nature of the SC to PG transition is described. A
new order parameter $\beta(T)$ naturally emerges, whose properties
are examined as a function of temperature, carrier concentration and
disorder - i.e. the phase diagrams. It extends the notion of `pair
field' that we reported in the coarse-grain approach \cite{smn}, and
gives a direct link between the SC coherence and the parameters of
the QP spectral function. \\

While the Bose-like condensation of preformed pairs is relatively
straightforward, the interpretation of the experimental QP spectra
suggests a much larger number of parameters. In this work we argue
that the further complexity is due to the {coupling of
quasiparticles to Cooper-pair excitations}. This proposition is
strongly supported by the variety of experimental spectra, analysed
in sections IV and V, and other widely accepted properties of
high-$T_c$.

\section{II.\,Hamiltonian for interacting pairs}\label{hamiltonian}

In the absence of pre-formed pairs, the so-called `pairing'
Hamiltonian \cite{tinkham} takes the form\,:
\begin{equation}\label{hamA}
H = H_0 + H_{pair} + H^{'}
\end{equation}
where, in standard notation,
\begin{equation}\label{elec}
H_0 = \sum_{k,\sigma}\ \epsilon_k\ {c_{k,\sigma}}^\dag\ c_{k,\sigma}
\end{equation}
\begin{equation}\label{pair}
H_{pair} = -\sum_k\ (\Delta_k\,{b_k}^\dag + {\Delta_k}^*\,{b_k} -
\Delta_k\,<b_k^*>)
\end{equation}
with\,: ${b_k} = {c_{-k,\downarrow}}^\dag\ c_{k,\uparrow}$ and
$H^{'}$ is bilinear in the fluctuation terms\,: $b_k\,-\,<b_k>$.
Now, in the problem we are addressing, we consider pre-existing
pairs that we label by $i$. We assume that these pairs are still
described by the BCS model, so that eq.\,(\ref{pair}) takes the
form\,:
\begin{equation}\label{PGham}
H_{pair} = -\sum_i\,\sum_k\ (\Delta^i_k\,{b^i_k}^\dag +
{\Delta^i_k}^*\,{b^i_k})
\end{equation}
after dropping the constant term. We assume that the bilinear
fluctuating terms in $H^{'}$ that are diagonal in $i$ only
renormalize the pair amplitude. On the contrary, we keep the
non-diagonal terms such that a new {\it mutual interaction} between
pairs emerges. We find $H^{'} = H_{int}$ can be written\,:
\begin{equation}\label{Hint}
H_{int} = \frac{1}{2}\, \sum_{i,j} \sum_{k,k'}\, \beta^{i,j}_{k,k'}\
{b^j_{k'}}\,{b^i_{k}}^\dag + h.c.
\end{equation}
where $\beta^{i,j}_{k,k'}$ is the microscopic coupling. This term is
neglected in the conventional approach.

In the absence of the pair-pair interaction $H_{int}=0$, then
$H_{PG} = H_0 + H_{pair}$ describes a {\it pseudogap state} of
uncondensed pairs having a binding energy compatible with
$\sim\,T^*\gg\,T_c$. As expressed in eq.\,(\ref{PGham}), they have a
fluctuating amplitude $\Delta^i_k$ distributed according to\,:
\begin{equation}\label{lorentz}
P_0(\Delta^i_k) = \frac{\sigma_0^2}{(\Delta^i_k-\Delta_{0,k})^2 +
\sigma_0^2}
\end{equation}
where $\Delta_{0,k}$ and $\sigma_0$ are the average pairing energy
and width, respectively. These two parameters will remain central to
our model throughout this work.

In the non superconducting state, the excitation spectrum is\,:
$E^i_k = \sqrt{\epsilon_k^2 + |\Delta^i_k|^2}$ with the
corresponding spectral function\,:
\begin{equation}\label{specfun}
{A}(\epsilon_k, E) = -\frac{1}{\pi}\ {\rm Im}\ \int_0^\infty
d\Delta^i_k\ P_0(\Delta^i_k) \frac{1}{E - E^i_k + i\,\Gamma}
\end{equation}
where $\Gamma$ is the Dynes QP broadening parameter \cite{dynes}. As
shown in \cite{smn}, the above spectral function and corresponding
density of states (DOS) leads to a smeared gap of width $\Delta_0$
at the Fermi level characterized by {\it vanishing coherence peaks}
(similar to Fig.\,\ref{f4}, spectrum 5).

The interaction between Cooper pairs, $\beta^{i,j}_{k,k'}$, is
responsible for the superconducting state. As in our previous work
\cite{smn}, this interaction is of the form\,:
\begin{equation}\label{betaij}
{\beta^{i,j}_{k,k'}} =
g_k\,g_{k'}\,P_0(\Delta^i_k)\,P_0(\Delta^j_{k'})
\end{equation}
where the factor $g_k$ preserves the $d$-wave symmetry, and
$P_0(\Delta^i_k)$ is the pair distribution. With this potential the
equalizing of the $\Delta^i_k$ in the final SC state is favored
while retaining the memory of the initial state. It has the useful
property of being separable, which allows for the decoupling of the
equations.

In the mean-field approximation, the operator $b^j_{k'}$ is replaced
by its quantum average\,: $<b^j_{k'}> =
\Delta^j_{k'}/(2\,E^j_{k'})$, so that the effective interaction
reduces to\,:
\begin{eqnarray}\label{Hint2}
H_{int} = \sum_{i, k} 2\,\beta_k\, P_0(\Delta^i_k)\,{b^i_{k}}^\dag +
h.c. &\null&\\
{\rm where,}\ \ \ \ \ \ \ \ \beta_k = g_{k}\,\sum_{j, k'}\ g_{k'}\
P_0(\Delta^j_{k'})\
\frac{\Delta^j_{k'}}{8\,E^j_{k'}}&\null&\label{Hint2B}
\end{eqnarray}
which defines $\beta_k$ in terms of all other pairs $j \neq i$ in
the system.

We assume that in the PG state, $\beta_k$ is negligible due to the
fluctuations of $\Delta^j_k$, see eq.\,(\ref{Hint2B}). On the
contrary, upon condensation, $\Delta^j_k\rightarrow \Delta_{k}$, one
has\,:
\begin{equation}\label{betasc}
\beta_k = N_p \times g_{k}\,\sum_{k'}\ g_{k'}\ P_0(\Delta_{k'})\
\frac{\Delta_{k'}}{8\,E_{k'}}
\end{equation}
which is much larger due to the factor $N_p$, the number of pairs.
Moreover, eq.\,(\ref{betasc}) is a new self-consistent equation
which depends on the exact $\Delta_{k}$ and the corresponding QP
excitations \,: $$E_k = \sqrt{\epsilon_k^2 + \Delta_k^2}$$

To obtain the SC gap equation, we first assume that all pairs join
the condensate (a condition we relax in Sec.\,III)\,:
${b^i_{k}}^\dag \rightarrow {b_{k}}^\dag$. Then $H_{pair}$ takes the
form\,:
\begin{equation}\label{Hpair}
H_{pair} = - \sum_{k}\ (\Delta_{0,k}\,{b_k}^\dag +
{\Delta_{0,k}}^*\,{b_k})
\end{equation}
where $\Delta_{0,k}$ is average value in the initial state
($\beta_k=0$).

The interaction term $H_{int}$ is evaluated directly in the final
state ($\Delta^i_{k} \rightarrow \Delta_{k}$), in the spirit of the
Brillouin-Wigner perturbation approach\,:
\begin{equation}\label{impair}
H_{int} = 2\,\sum_{k} \beta_k\, P_0(\Delta_k)\,{b_{k}}^\dag + h.c.
\end{equation}
Combining (\ref{Hpair}) and (\ref{impair}), the total Hamiltonian is
then\,:
\begin{equation}\label{Hsc}
H = H_0 - \sum_{k}\ (\Delta_{k}\,{b_k}^\dag + {\Delta_{k}}^*\,{b_k})
\end{equation}
with the final gap equation\,:
\begin{equation}\label{gapeq1}
\Delta_{k} = \Delta_{0,k} - 2\, \beta_k\,P_0(\Delta_{k})
\end{equation}
Written in this form, the equivalence of the separable pair-pair
interaction (\ref{betaij}) to an additional `field' is evident.

In this work,  $\Delta_{k}$ is assumed to be $d$-wave and, at the
Fermi-level ($\epsilon_k = 0$) along the anti-nodal direction, we
simply note $\Delta_{k} = \Delta_{p}$\,:
\begin{equation}\label{gapeq2}
\Delta_{p} = \Delta_{0} - 2\,\beta\,P_0(\Delta_{p})
\end{equation}
We stress that the effect of the interaction (- $\beta$ above) is to
{\it lower} the pair binding energy with respect to $\Delta_{0}$.

\begin{figure}[t]
\centering
\includegraphics[width=7.0 cm]{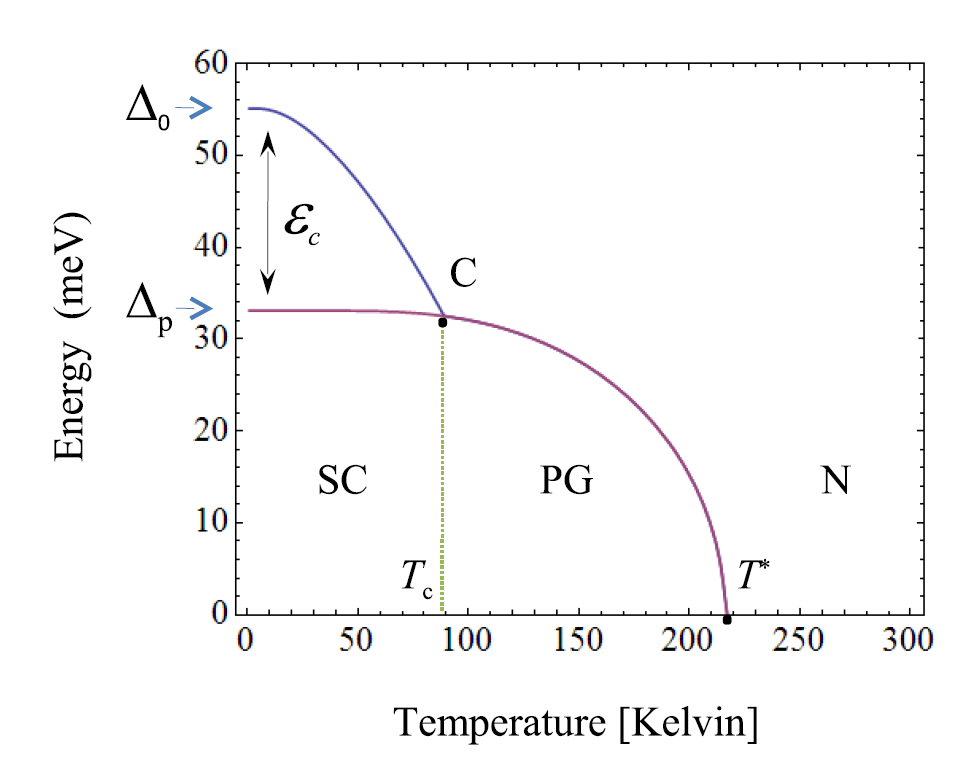} \caption{Phase
diagram in the interacting-pair model.}\label{f1}
\end{figure}

The maximum (anti-nodal) gap $\Delta_{p}$ appears on both sides of
eq.~(\ref{gapeq2}) which must be solved self-consistently. Defining
the change in energy at $T = 0$ as $$\varepsilon_c  = \Delta_{0} -
\Delta_{p}$$ we obtain the cubic equation in $\varepsilon_c$\,:
\begin{equation}\label{econ1}
\varepsilon_c = 2\,\beta\ P_0(\Delta_p) = 2\,\beta
\frac{\sigma_0^2}{\varepsilon_c^2 + \sigma_0^2}
\end{equation}
which shows that $\varepsilon_c$ has a strong dependence on the
interaction $\beta$ as well as the distribution width, $\sigma_0$.
For optimally-doped BiSrCaCuO(2212), $\varepsilon_c \simeq$ 22 meV,
which is still large compared to $k_B\,T_c$, but less than the
spectral gap $\Delta_{p} \simeq$ 38 meV. The three characteristic
energies\,: $\Delta_{p}$, $\varepsilon_c$, and $\Delta_{0} =
\Delta_{p} + \varepsilon_c$, are illustrated in the phase diagram
Fig.~\ref{f1}.

The interaction $\beta$ has a dominant effect on the energy changes
involved in the transition. To illustrate, consider the BCS
expression for the self-consistent gap\,: $$\Delta_p = 2\,E_0\, {\rm
e}^{-1/(N_0\,V_p)}$$ where $E_0$ is the cut-off energy, $N_0$ is the
Fermi-level DOS and $V_p$ is the {\it total} pair potential. Using
$V_p = V_0 + V_{int} = V_0 (1-\overline \beta)$, where $\overline
\beta$ is the relative pair-pair interaction, and $\Delta_0$ and
$\varepsilon_c$ are of the form\,:
\begin{equation}\label{betabar}
\Delta_0 = \Delta_p\,(1+\overline \beta) \ \ \ \ {\rm and } \ \ \
\varepsilon_c = \overline \beta\,\Delta_p
\end{equation}

\begin{figure}[h!]
\vbox to 6.2 cm{ \centering
\includegraphics[width=5.8 cm]{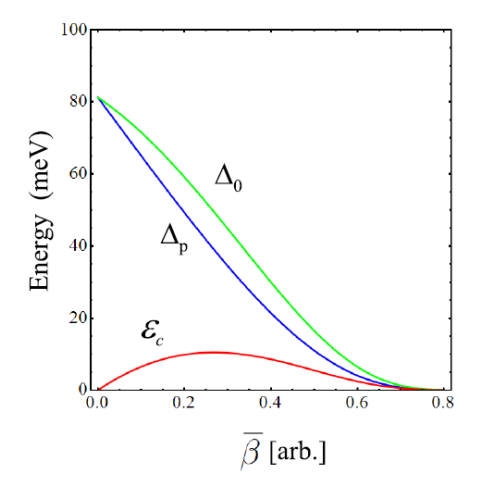} \caption{Influence
of a pair-pair interaction, $\overline \beta$}\label{f2}}
\end{figure}

As illustrated in Fig.~\ref{f2}, the spectral gap $\Delta_p$
decreases sharply as a function of $\overline \beta$, becoming
vanishingly small as $\overline \beta \rightarrow 1$. $\Delta_0$
begins equal to $\Delta_p$, for $\overline \beta = 0 $, and then
follows a similar trend as $\Delta_p$, albeit with $\Delta_0 >
\Delta_p$. The main observation is that $\varepsilon_c$ is nearly
dome-shaped -- a general property of high-$T_c$ and indeed other
types of superconductors. Here, the pair-pair interaction $\beta$
causes the rapid decrease in $\Delta_p$ and the dome-shape of the
condensation energy.

The physical origin of the pair-pair interaction within the
condensate can now be addressed. We consider that it is a {\it
non-retarded interaction mediated directly by the quasiparticles}.
Consequently it is long-range, implying that $\beta$ is proportional
to the number of condensed pairs $N_{oc}(T)$. It can therefore be
written\,:
\begin{equation}\label{beta}
\beta(T) = \beta_0\,N_{oc}(T)
\end{equation}
The effect of this interaction on the condensation mechanism is
investigated in the following Sec.\,III and on the quasiparticle
spectrum in Sec.\,IV.

\section{III.\,Pair condensation}\label{pairs}
Consider the phase diagram of Fig.~\ref{f1} in the context of
uncondensed pairs above $T_c$. Qualitatively, the total pair energy
$\Delta_p$ follows the critical curve shown and which vanishes at
$\sim T^*$. In particular, the SC critical temperature, denoted by
{\bf C}, is situated on the {\it plateau} of the critical curve
where $\Delta_p$ is about equal to its zero-temperature value. The
fact that $\Delta_p$ remains constant in the SC to PG transition can
be directly seen as a function of temperature
\cite{rennerT,matsuda}, disorder \cite{crendisorder,mcelroy} and in
the vortex core \cite{rennerB,phd}.

In the BCS gap equation, the reduction of the gap with rising
temperature below $T_c$ is associated with quasiparticle excitations
given by the factor \,: $$1-2f(E_k,T) = {\rm
tanh}\left(\frac{E_k}{2\,k_B\,T}\right)$$ where $f(E_k,T)$ is the
Fermi-Dirac function and $E_k$ the QP energy. In the case of
cuprates, as the temperature approaches $T_c$ from below, the
smallest value of this function is ${\rm
tanh}(\Delta_p/(2\,k_B\,T_c)) \sim .96$ and is $\sim 1$ otherwise.
Clearly, pair-breaking linked to thermally excited quasiparticles is
not relevant.

Assuming 2D free-electrons for a single conducting plane, we note
that the pair density per unit area is about\,: $N_p/A \simeq
m\Delta_p/(2 \pi \hbar^2)$. Then, consider the typical number of
pairs within the disk of area $S = \pi \xi^2$, with $\sim \xi$ being
a pair diameter\,:
$$N_p \simeq \frac{m\Delta_p\, \xi^2}{2 \hbar^2} \sim 1-2$$ Clearly,
the combination of small $\xi$ and small pair overlap strongly
supports an interacting boson model.

\vskip 2mm

\begin{table}[h!]
\caption{Summary of main parameters}\label{parameters} \vskip 0 mm
\begin{tabular}{@{}cccll@{}}
\hline\hline\toprule \multirow{4}{*}{SC gap equation} & $\Delta_p$ &
\multicolumn{3}{l}{self-consistent SC gap} \\ \cmidrule(l){2-5}
 & $\beta$ & \multicolumn{3}{l}{pair-pair interaction} \\ \cmidrule(l){2-5}
 & $\Delta_0$ & \multicolumn{3}{l}{pair distribution average} \\ \cmidrule(l){2-5}
 & $\sigma_0$ & \multicolumn{3}{l}{pair distribution width} \\ \midrule
Condensate & $\delta$ & \multicolumn{3}{l}{excitation mini-gap} \\
\midrule \multirow{2}{*}{QP spectrum} & $\sigma$ &
\multicolumn{3}{l}{pair broadening} \\ \cmidrule(l){2-5}
 & \multicolumn{1}{c}{$\Gamma$} & \multicolumn{3}{l}{lifetime broadening} \\ \bottomrule
\end{tabular}
\end{table}

\begin{figure*}
\vbox to 9.4 cm{ \centering
\includegraphics[width=11.8 cm]{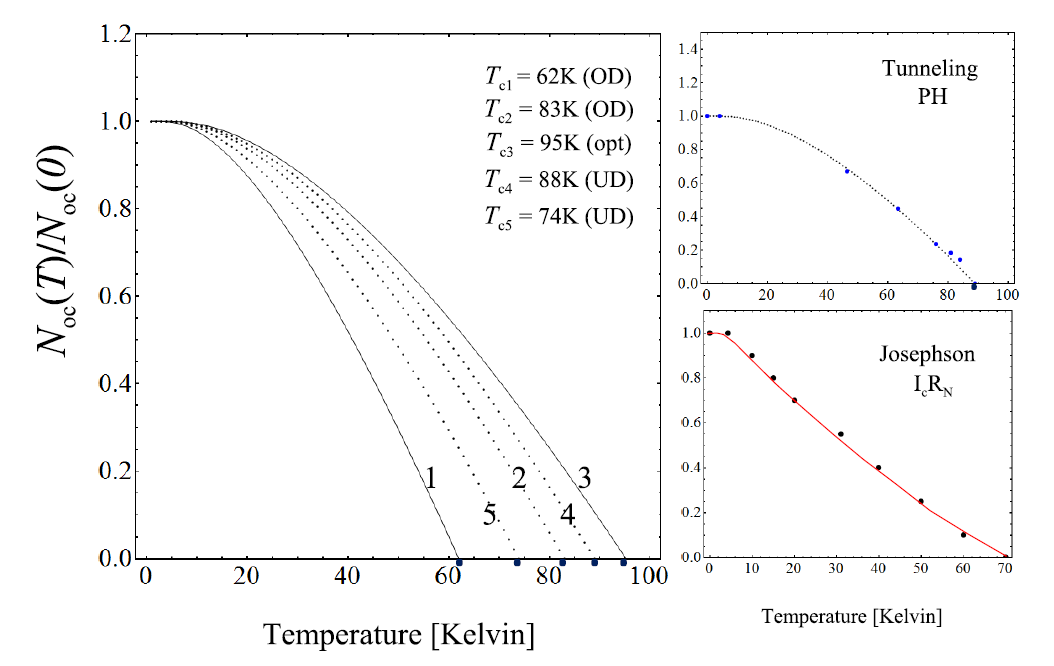}
\captionsetup{format=hang} \caption[]{\small Temperature dependance
of the Bose condensate
pair density $N_{oc}(T)$ using eq.~(\ref{bose2}). \\
Upper right\,: Fit to QP DOS peak height from \cite{rennerT} (reported $T_c$ = 83 K)\\
Lower right\,: Josephson $I_c R_N$ from SIS break-junction \cite{miyakawa} (reported $T_c$ = 77 K)\\
Left panel\,: $N_{oc}(T)$ corresponding to 5 samples from overdoped
(1) to underdoped sides (5).} \label{f3}
\parindent 0 cm
}
\end{figure*}

We thus retain the preformed-pair hypothesis above $T_c$ and propose
that their condensation follows Bose statistics. To proceed, one
needs the excited states of the system which, in our model, are
given by the previous distribution $P_0(\Delta^i)$. In this Section,
the magnitude of the gap is relevant to these excited pairs and we
drop the explicit $k$-dependence.

The occupation number, $N_{oc}(T)$, vanishes in the
non-superconducting PG state, $N_{oc}(T_c)=0$, all pairs being in
excited states. On the contrary, $N_{oc}(0) = 1$ in the SC state at
zero-temperature, at which all pairs have the unique value\,:
$\Delta^i = \Delta_p$. Analytically, the model would then imply\,:
\begin{equation}\label{bose1}
N_{oc}(T) = \mathcal{A}\,\int_{\Delta_p}^{\infty}\,{\rm d}\Delta^i\
P_0(\Delta^i)\ f_B(\Delta^i-\Delta_p, T)
\end{equation}
$f_B(\Delta^i, T)$ is the Bose distribution, with vanishing chemical
potential ($\mu_B\rightarrow 0$), $\Delta^i-\Delta_p$ is the pair
excitation energy and $\mathcal{A}$ is for normalization. It is
evident that this integral diverges since at the lower limit where
$\Delta^i \rightarrow \Delta_p$ the distribution $P_0(\Delta^i)$
remains finite while $f_B(\Delta^i, T)$ diverges. This well-known
singularity indicates that a 2D condensation of non-interacting
bosons cannot occur \cite{fetter}. Furthermore, a
Kosterlitz-Thouless transition can also be ruled out at this
temperature scale.

There must therefore be a strong {\it attractive interaction}
between a given boson and the condensate. Since the Bose
distribution in eq.~(\ref{bose1}) decreases sharply with energy, the
low-lying excitations just above $\Delta_p$ are critical. A simple
model for the energy needed to remove a pair from the condensate is
a mini-gap $\delta$ separating the condensate energy from the
excited states. We thus write\,:
\begin{equation}\label{bose2}
N_{oc}(T) = \mathcal{A}\,\int_{\Delta_p+\delta}^{\infty}\,{\rm
d}\Delta^i\ P_0(\Delta^i)\ f_B(\Delta^i-\Delta_p, T)
\end{equation}
to calculate the occupation number. The free-energy at the
transition can likewise be written \cite{fetter}\,:
\begin{equation}\label{bose3}
\mathcal F(T_c) = \mathcal{A}\,\int_{\Delta_p+\delta}^{\infty}\,{\rm
d}\Delta^i\ P_0(\Delta^i)\ g(\Delta^i-\Delta_p, T_c)
\end{equation}
where $g(\Delta^i, T) = - k_B\,T\, {\rm ln}(1 - {\rm
exp}(-\Delta^i/(k_B T))$.

The interpretation of the mini-gap $\delta$ will be discussed later.
In practise, we use the experimental value of $T_c$ to determine the
normalization $\mathcal{A}$ of the pair distribution. For an
optimally-doped BiSrCaCuO(2212), implementing eq.~(\ref{bose2}) with
$\delta = 2.1$ meV and $T_c = 95$ K leads to $\mathcal F(T_c) \simeq
0.84\, k_B\, T_c$, which is a satisfactory order of magnitude.

\begin{figure*}
\vbox to 10.8 cm{
\begin{center}
\includegraphics[width=13 cm]{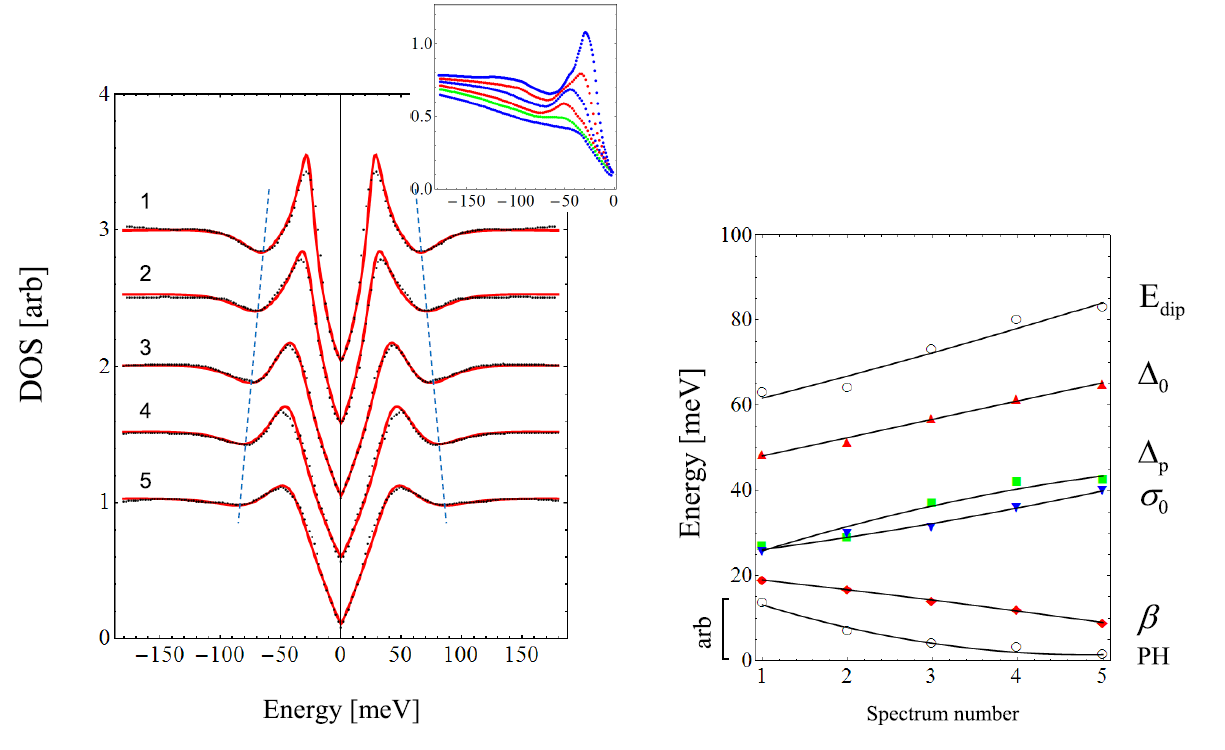}
\end{center}
\captionsetup{format=hang}\caption[]{\small Left panel\,: Fits to a
series of 5 QP spectra from McElroy et al.\cite{mcelroy} using
eq.~(\ref{finDOS}) (dots\,: experiment, solid line\,: theory). The
spectra are ordered according to the `strength' of their SC
features. Insert\,: Experimental spectra on the occupied side before
removing of background slope. \\
Right panel\,: Values of important parameters from the spectra.
$\Delta_0$, $\sigma_0$, and  $E_{dip}$ are all moving upwards,
following $\Delta_p$. The pair-pair interaction $\beta\propto
N_{oc}$ moves oppositely towards lower energy, parallel to the peak
heights (PH). \\ $\sigma$ varies from 2 to 16 meV from spectrum 1 to
5 (not shown).} \label{f4}}
\end{figure*}

The above Fig.\,\ref{f3} summarizes just how well the Bose
condensation model fits the data. First, precise fits to the QP
tunneling spectra from Renner et al.\cite{rennerT}, as a function of
$T$, allow to determine the parameters $\Delta_0$ and $\sigma_0$, as
well as the QP peak heights (PH). As we showed in our previous work
\cite{smn,sacks}, the attenuation of the DOS peak height, analogous
to the effect shown in the spectra of Fig.~\ref{f4}, is due to the
gradual decrease in the interaction $\beta(T)$, as expected from
eq.\,(\ref{beta}). The condensation energy $\varepsilon_c(T)$ is
also proportional $N_{oc}(T)$, eq.~(\ref{econ1}), due to its
dependance on $\beta(T)$.

In the upper right panel, we plot both the data points from Renner
et al.\cite{rennerT} and the best fit of $N_{oc}(T)$ using
eq.~(\ref{bose2}), which is excellent except for a slight jump near
$T_c$. Unfortunately, similar STS experiments as a function of
temperature are rare. In the lower right panel of Fig.\,\ref{f3} we
plot the critical current $I_c R_N$ obtained by Miyakawa et al.
using a SIS break-junction setup \cite{miyakawa}. We emphasize that
the Josephson $I_c R_N$ is directly sensitive to the SC condensate.
Here it decreases monotonically with $T$ whereas the gap value
$\Delta_p$ {\it remains constant} in this temperature range, in
agreement with our model. In both experiments, there is a shift in
$T_c$ comparing the local vanishing of $N_{oc}(T)$ and the reported
bulk value. Still, the quantity $\beta(T) \propto N_{oc}(T)$ has the
required properties for an order parameter.

The overall behavior of $N_{oc}(T)$ from underdoped to overdoped
BiSrCaCuO(2212) is illustrated in the main panel of Fig.\,\ref{f3}.
The curves are quite different from the conventional BCS order
parameter having a short plateau towards low-$T$ (spanning $\sim 10$
K) then descending rapidly to intersect the $T$-axis with a finite
slope. Starting from the fitted estimate of the mini-gap, $\delta =
2$ meV, we scaled $\delta$ with $T_c$ using $3.85\,\delta = k_B\,
T_c$. The values of the parameters $\Delta_0$ and $\sigma_0$ of the
pair distribution were obtained from precise fits to the
experimental spectra on BiSrCaCuO(2212), the subject of the
following section.

\section{IV.\,Quasiparticle spectrum}\label{qpspec}

The QP spectral function and DOS, obtained by ARPES and tunneling
respectively, contain valuable information on the SC state at the
microscopic scale. The SC spectral function is\,:
\begin{equation}
{A}(\epsilon_k, E) = -\frac{1}{\pi}\ {\rm Im}\ \frac{1}{E - E_k +
i\,\Gamma}
\end{equation}
where $E_k$ is the QP energy. The BCS-Bogoliobov coherence factors
are omitted since, as described by Schrieffer \cite{schrieffer},
they are redundant if $\Delta_k(\pm \epsilon_k)$ is symmetric with
respect to the Fermi level.

The QP DOS is obtained directly from the spectral function using\,:
$N_S(E) = 2\,\sum_k\ A(\epsilon_k, E)$ where, for a $d$-wave pairing
symmetry, $\Delta_k(\theta) = \Delta_p\,{\rm cos}(2\theta)$ leading
to the typical V-shaped DOS \cite{mallet,hoogenboom,fischer_rev}.
However, as we have shown repeatedly \cite{constraints,sacks,smn},
such a $d$-wave DOS, with weak logarithmic singularities at the
coherence peaks and no dip-hump structures, fails to match the
experimental spectra.

We have shown that the total Hamiltonian (\ref{hamA}) with the
interaction term (\ref{Hint2}) not only expresses the ground-state,
with all $\Delta^i_k$ within the condensate $\Delta^i_k = \Delta_k$,
but also the {\it pair} excited states, with the corresponding
distribution, $P_0(\Delta^i_k)$. The latter excited states lower the
condensate occupation number $N_{oc}(T)$, eq.\,(\ref{bose2}), and
eventually the SC coherence is lost at $T_c$. How does this
relatively simple Bose-like transition affect the QP spectral
function\,?

In our model, the pair-pair interaction is directly mediated by the
quasiparticles. An excited pair, i.e. $\Delta^i_k
> \Delta_k$, can be coupled to a quasiparticle of the same energy and
conversely, that an excited QP of energy $E^i_k$ is linked to such a
`virtual' pair. We propose that this energy-conserving coupling is
manifested directly in the pair-pair interaction term of the
Hamiltonian, eq.~(\ref{Hint2}). Indeed, replacing $\Delta^i_k$ by
$E^i_k$ we have\,:
\begin{equation}\label{Hint3}
H_{int} = \sum_{i, k}\ 2\, \beta_k\,P_0(E^i_k) \,{b^i_{k}}^\dag +
h.c.
\end{equation}
In the final SC state, this leads to a non-retarded energy-dependant
gap\,:
\begin{equation}\label{Egapfun}
\Delta_k(E_k) = \Delta_{0,k} - 2\, \beta_k\,P_0(E_k)
\end{equation}
Note that the QP dispersion, while having the same functional
form\,:
\begin{equation}\label{dispersion}
E_k = \sqrt{\epsilon_k^2 + \Delta_k(E_k)^2}
\end{equation}
is in fact greatly modified due to this energy dependence and must
be used for calculating the corresponding DOS. With surprisingly few
parameters, this QP dispersion and gap function
eqs.~(\ref{Egapfun},\ref{dispersion}) accurately matches the wide
peaks and characteristic dips as measured in the spectra, see
Fig.\ref{f4}.

The derivation of the DOS is given a full treatment in
\cite{constraints,smn}. We recall that, neglecting
$\Gamma$-broadening, it depends on the following derivative\,:
\begin{equation}
N_S(E,\theta) = \frac{N_0}{2\,\pi}\ \int_0^\infty d\epsilon_k\
\delta(E_k - E) = \frac{N_0}{2\,\pi}\
\left[\frac{\partial\epsilon_k}{\partial E_k}\right]_{E_k=E}
\end{equation}
where $N_S(E,\theta)$ is the partial DOS in the $\theta$ direction.
The final DOS therefore involves the energy-derivative of the gap
function itself - a rather unusual property. A full treatment, using
the spectral function, leads to\,:
\begin{equation}\label{finDOS}
\hskip -2 mm N_S(E) = N_0\ {\rm Re} \int_0^{2\,\pi}
\frac{d\theta}{{2\,\pi}} \ \frac{E - i \Gamma - \Delta_k(E,\theta)
\frac{\partial\Delta_k(E,\theta)}{\partial E}}{\sqrt{(E - i
\Gamma)^2 - \Delta_k(E,\theta)^2}}
\end{equation}
This unique DOS function is used to fit all the spectra in this
work, Figs.\ref{f4} and \ref{f5}, where the spectral shape is
determined by the distribution $P_0(E_k)$, i.e. $\Delta_0$ and
$\sigma_0$, and the value of $\beta$ (see Table \ref{parameters}).

Above $T_c$ there is the additional effect of thermal pair
excitations above $\Delta_p$. Also, pair excitations arise due to
disorder, which we take into account by the broadening parameter
$\sigma$ of the pair potential\,: $\Delta_p \rightarrow \Delta_p-i\,
\sigma$ in eq.\,(\ref{finDOS}). Its value is important when
$N_{oc}(T) \ll 1$ or at low-doping (as shown in Fig.~\ref{f6}).
Otherwise, in all fits, $\Gamma \sim 1.5$ meV, is small.

In the detailed experiments by McElroy et al. \cite{mcelroy} full
STS mappings of BiSrCaCuO(2212) have been performed to study the
local SC characteristics and, with atomic resolution, the effects of
the local oxygen dopants. The study of three different samples with
nominal average gaps (45, 55 and 65 meV) are of particular interest,
ranging from near optimal to increasingly under-doped. The DOS
variation due to atomic oxygen dopants is quite significant\,: a
resonance near -.96 meV is seen on the occupied side of the spectrum
and leads to a strong variation of the SC spectral shape.

Two other points in these experiments are significant for the
present model. First, local spectra from one sample can be
identified with spectra from another sample at locations having
common characteristics (gap width, peak height, dip position).
Although the samples show topological variations, this suggests that
on the mesoscopic scale, the SC gap function has consistent
properties {\it independent of the sample} and its parameters are
physically meaningful. Secondly, the optimally-doped sample is much
more homogeneous than the other two - suggesting that a percolation
effect must be taken into account for very underdoped samples.

The spectra in Fig.\ref{f4} are sequenced according to the
`strength' of their superconducting characteristics\,: in spectrum
1., a smaller gap width, with sharper QP peaks and in spectrum 5., a
larger gap and very attenuated peaks. The spectra with strong SC
features largely dominate on the optimally-doped sample, whereas the
weak SC spectra are found in mesoscopic regions in the inhomogeneous
under-doped sample. The series resembles the SC to PG transition at
low-temperature as seen by Renner and Cren
\cite{rennerT,crendisorder} and in this context we consider the
compatibility of our model.

In Fig.\ref{f4}, left panel, we show the fits to the spectra, having
removed the background slope and symmetrizing in each case. In the
right panel we plot the values of the parameters in the DOS function
eq.~(\ref{finDOS}) and their variation from spectrum to spectrum.

The first point is that our gap value $\Delta_p$ is 10-20 \% smaller
than the nominal ones cited by McElroy et al. Indeed, the observed
spectral gap is affected by the derivative of the gap function at
$E_k \simeq \Delta_p$, as illustrated in Fig.\ref{f5} lower panel.
The measured peak is also affected by the parameter $\sigma$, which
reaches $\sim$ 16 meV for spectrum 5. This damping effect is smaller
in the case of sharp QP spectra.

Secondly, it is remarkable how well several parameters scale with
the spectral gap, in particular $\Delta_0$, $\sigma_0$ and the dip
position $E_{dip}$. On the contrary, the pair interaction parameter
$\beta$ moves in the {\it opposite direction}, concomitant with the
lowering of spectral peak heights (PH).

This observed increase of gap width $\Delta_p$ with decreasing
$\beta$ is a general consequence of the pair-pair interaction model,
eq.~(\ref{gapeq2}). Moreover, the discussion on the temperature
dependance of $\beta$, where we have $\beta(T) \propto N_{oc}(T)$,
leads to the conclusion that the condensate occupation density
$N_{oc}$ is decreasing from spectrum 1 to 5. This is confirmed by
the quantity $\varepsilon_c = \Delta_0 - \Delta_p$, which remains
relatively robust in this sequence, and $\sigma_0$ which becomes
large. The likely scenario is the effects of disorder in the
underdoped regime \cite{hirschfeld,ghosal}.

In summary, the simple QP spectrum, eq.\,(\ref{finDOS}), matches
remarkably well the measured local DOS of an inhomogeneous SC state,
e.g. spectra 3-5. In a given region of the sample, the main effect
is thus the change in magnitude of the condensate density
$N_{oc}(T)$ on the mesoscopic scale and, through the pair-pair
interaction $\beta$, the damping effect on the QP features
(peak-height and dip strength). Lastly, spectrum 5 shows very weak
quasiparticle peaks, close to the PG state measured at low
temperature by \cite{rennerT,crendisorder,phd}. In our view, the
{\it bona fide} PG state corresponds to the strict vanishing of both
$\beta$ and $\delta$, the two parameters linked to the condensate.


\section{V.\,Energy phase diagram}\label{doping}

\begin{figure}[h!]
\begin{center}
\includegraphics[width=8.8 cm]{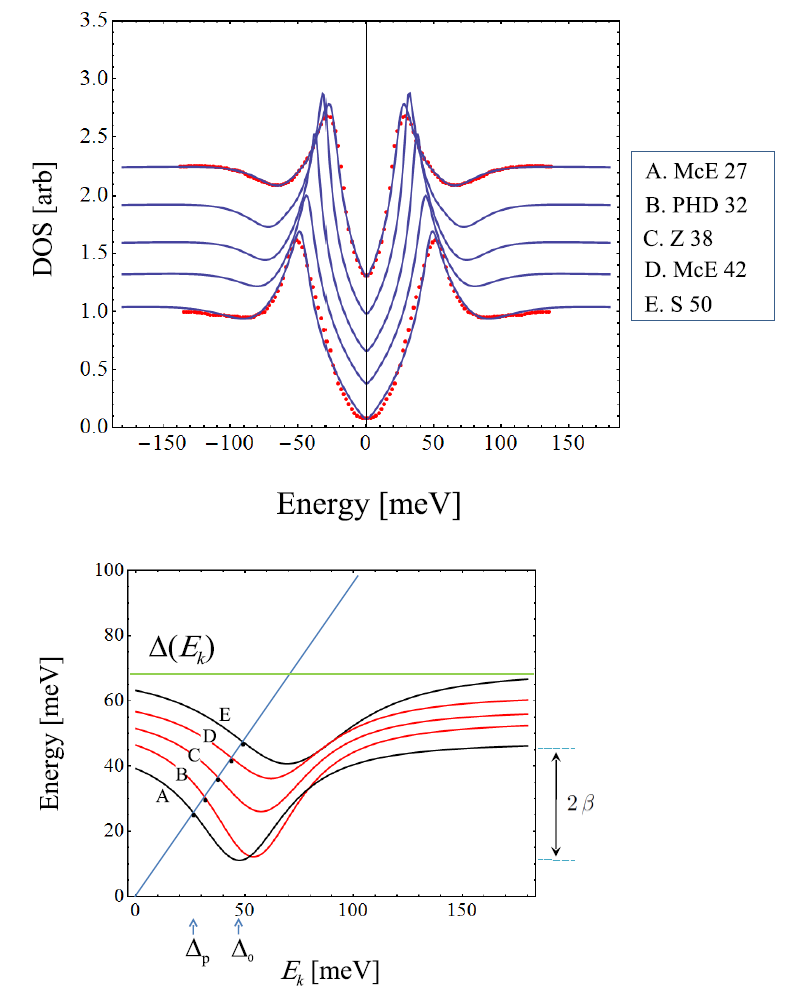}
\end{center}
\caption[]{\small Upper panel\,: QP DOS using eq.~(\ref{finDOS}) to
fit a series of 5 spectra A-E from various authors (A\,:
\cite{mcelroy}, B\,: \cite{phd}, C\,: \cite{miyakawa}, D\,:
\cite{mcelroy}, E\,: \cite{sugimoto}) ordered according to gap
width\,: A (overdoped), C (optimal), E (underdoped). (Data only
shown in A and E.) Lower panel\,: Plot of the corresponding
gap-functions $\Delta_k(E_k)$ A-E with fitted parameters\,: gap
width $\Delta_p$, minimum amplitude $\Delta_0$, and distribution
width, $\sigma_0$. These parameters are plotted in the $T-p$
diagram, Fig.~\ref{f6}.}\label{f5}
\end{figure}

We have chosen a wide variety of tunneling DOS from the literature
\cite{sugimoto,rennerT,mcelroy,phd,constraints,miyakawa,zazaSIS,fang}
to study the energy scales, their doping dependence and the possible
role of the pair-pair interactions. Five of these spectra labeled
A-E in Fig.~\ref{f5}, have been selected for their regularity and
the apparent stability of their parameters. In the upper panel, the
QP DOS, using eq.~(\ref{finDOS}), corresponds to underdoped, with
larger $\Delta_p\simeq 50$ meV, to overdoped, with smallest
$\Delta_p\simeq 27$ meV.

In the lower panel, Fig.~\ref{f5}, we show the gap functions
$\Delta_k(E_k)$ used to precisely fit each spectrum of the upper
panel. The experimental data are only shown for spectra A and E for
clarity and, as in Fig.~\ref{f4}, only the SC part of the spectrum
is considered. The quasiparticle peaks in the DOS are fairly sharp,
reaching $\sim 1.5-2$ above the background level. These are
determined by the negative slope of the gap function
${\partial\Delta_k(E_k)}/{\partial E_k}$ at $E_k = \Delta_p$. Higher
QP peaks have sometimes been measured, in particular
Ref.\,\cite{fang}, in good agreement with our model for large
$\beta$ and smaller $\sigma_0$.

Fig.~\ref{f5} reveals the regularity of the dip position, given by
the {\it positive} derivative of the gap-function near $\sim
2\,\Delta_p$, which increases from the overdoped to underdoped
sides. As in Refs.\,\cite{constraints, sacks, smn}, these
characteristics are found without a dynamical collective mode. It
remains to be investigated whether similar features seen in the
pnictide LiFeAs \cite{chi} have the same origin.

\begin{figure}[h!]
\centering
\includegraphics[width=8.4 cm]{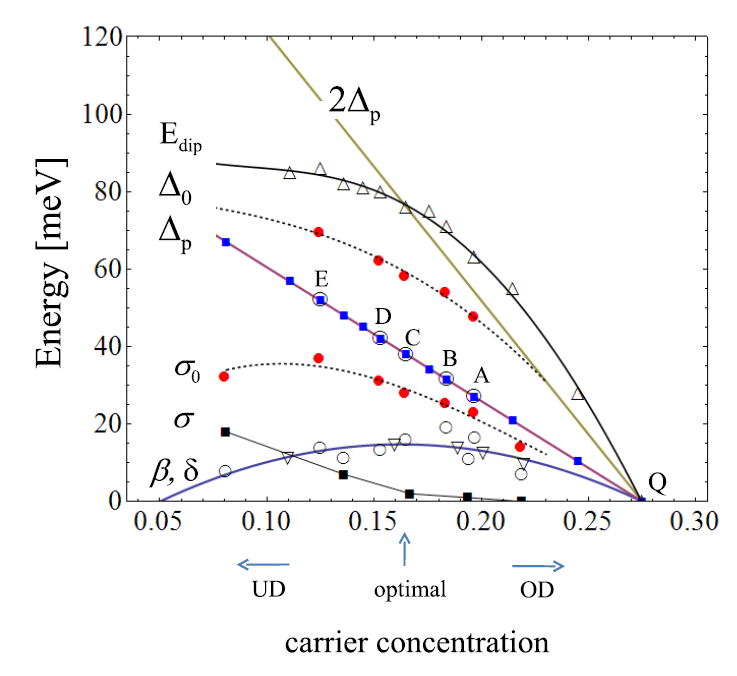}
\caption{\small Energy - doping ($p$) plot summarizing the
dependence of the parameters in the present model. Points A-E
correspond to the spectra of Fig.~\ref{f5}. The self-consistent gap
 $\Delta_p$ is linear; $\Delta_0$, $\sigma_0$ and $E_{dip}$
 are decreasing with $p$ but are slightly convex. At optimal doping
 $\varepsilon_c$, the pair-pair interaction
 $\beta$ and the mini-gap $\delta$ are maximal.
 Both $\beta$ and $\sim 7\,\delta$ follow the dome-shaped
 curve\,: $1.8\, k_B\,Tc$.}\label{f6}
\end{figure}

The parameters of these five spectra are then plotted in the $T-p$
phase diagram, Fig.~\ref{f6}. The self-consistent gap $\Delta_p$
follows a near-perfect linear trend decreasing from underdoped to
overdoped sides and extrapolating to zero at {\bf Q}. To serve as a
guide, the dome-shaped plot $1.8\,k_B\,T_c$ is shown. To a first
approximation, for the doping near the top of the $T_c$-dome, the
parameters $\Delta_0$, $\sigma_0$, and  $E_{dip}$ are remarkably
continuous and almost linear. This supports the conclusion that,
excepting the highly-underdoped samples, the parameters deduced from
the spectra of Fig.~\ref{f5} are reliable.

A closer look reveals the convex shape of $\Delta_0$, with the
consequence on $\varepsilon_c = \Delta_0 - \Delta_p$, as well as
$E_{dip}$, also being convex. The dip position \cite{sacks,zazaSIS}
evolves from below $2\,\Delta_p$ to above $2\,\Delta_p$ for
increasing $p$, as resolved by many tunneling spectra. On the
right-hand side, all curves seem to converge to a single point {\bf
Q}.

Of major significance to the present model, the pair-pair
interaction follows $T_c$ along with the attractive mini-gap
$\delta$, as plotted in the lower part of Fig.~\ref{f6}. We find
$\beta\simeq 1.8\, k_B\,T_c$ and $\delta \simeq .26\, k_B\,T_c$ with
$\beta/\delta \simeq 7$. Admittedly,  $\beta$ reveals some scatter,
sometimes reaching $\sim 2.2\, k_B\,T_c$. We note the loss of
precision in the fits on the far under-doped side, where the
coherence peaks are attenuated. In this regime where $\beta$
decreases sharply, the number of excited pairs, related to $\sigma$,
increases markedly, so that $\sigma$ practically joins $\sigma_0$.
The available data indicates that, at this extreme left-hand side of
the $p$-diagram, both $\beta$ and $\delta$ are vanishing.

In our view, the overall scenario is as follows. A pair-pair
interaction, governed by the coupling $\beta_{i,j}$ allows for
incoherent pairs, with distribution $P_0(\Delta^i)$ to begin
aligning their energies, ultimately to the amplitude $\Delta_p$.
Since the interaction is mediated by the quasiparticles, it provides
the mechanism for establishing long-range order. The final pair-pair
interaction in the SC state, being proportional to the condensate
density, $\beta(T)= \beta_0\,N_{oc}(T)$, supports this hypothesis.
At low-doping, the coherent state cannot be formed if the distance
between pre-formed pairs is too large to be efficiently coupled,
since the spatial inhomogeneity that breaks the coherence will
result in a cut-off in the interaction. With increased doping, the
interaction favors long-range order, the system becomes more
homogeneous, and $T_c$ increases. At the same time, the pair-pair
interaction weakens the pair self-energy $\Delta_p$ (see
fig.\,\ref{f6}), which decreases uniformly from under-doped to
over-doped sides. Thus the SC state becomes weakened again.

In the context of pre-formed pairs, the question of a non-BCS
condensation mechanism arises. Indeed, in a 2D system, a
conventional Bose condensation is not possible without an
inter-particle interaction. We proposed that a mini-gap $\delta$ in
the {\it pair} excitation spectrum allows for the condensation. The
calculated condensate density $N_{oc}(T)$ using this model follows
very nicely the experimental data obtained both by single particle
tunneling and by Josephson effect (fig.\,\ref{f3}). The temperature
dependence thus has the properties of an order parameter; at the
critical temperature, $N_{oc}(T_c) = 0$ and we get back the
incoherent pairs of the PG state. The mini-gap $\delta$ is the
direct measure of the stability of the coherent state and, just like
$\beta$, it follows the $T_c$ versus $p$ curve.

Since $\delta$ is the energy to excite a single pair with respect to
the condensate, it can also be viewed as the energy needed to create
a local defect in the phase. Such a defect might be the rotation of
a local pair wave function with respect to the $d$-wave SC
condensate. Without the mini-gap, there is no condensation at all
and no coherent pair-pair interaction either ($\beta = 0$) -- the
two being intimately linked.

In contrast to $\delta$, which is the excitation energy of one pair,
$\varepsilon_c = \Delta_0-\Delta_p$ represents the adiabatic
condensation energy of all pairs from a PG state at zero-temperature
to the SC state. This energy is of the same order of magnitude as
$\beta$ to which it is proportional, see eq.\,(\ref{econ1}), and
therefore it too follows $T_c$. The quantity $\varepsilon_c$ also
reveals the unconventional shift in pair binding-energy\,: from
$\Delta_0$, the average value in the ($T=0$) PG state, to the
smaller value $\Delta_p$ in the SC state. This suggests that the
pairs give up potential energy when the system acquires long-range
order via the quasiparticles. This mechanism, important at low
doping, becomes increasingly inefficient as doping becomes large and
$\varepsilon_c$ is correspondingly small. Indeed, in the over-doped
regime the behavior is dominated by the weakening of the
superconductivity associated with the decrease of $\Delta_p$ with
$p$. Ultimately, all parameters vanish at a unique critical point
{\bf Q}.

The optimal doping appears at the best compromise between the
depairing associated with the decrease of $\Delta_p$, and the
increase of the homogeneity and pair concentration that stabilise
the coherent state -- both effects being driven by the pair-pair
interaction.

\section{VI.\,Conclusion}
In this work we have shown that the phase transition to
superconductivity from a pseudogap state of preformed pairs can be
understood in relatively simple terms and few parameters. It
consists of a Bose-like condensation with a small mini-gap $\delta$
above $\Delta_p$ which represents the minimum energy to remove a
pair from the condensate.

The low-level pair excitations {\it above $\Delta_p$} are thus
critical for the condensation mechanism. No thermal QP excitations,
\`{a} la BCS, are needed\,; only the pair excitation distribution
$P_0(\Delta^i_k)$ is involved. With this mechanism, the properties
of the condensate occupation number $N_{oc}(T)$ are in good
agreement with available experiments. Moreover, the Fermi-level
spectral gap $\Delta_p$ remains constant in this temperature range.

The problem is rendered complex by two additional properties of the
pair-pair interaction\,: $\sim 2\,\beta(T)\,P_0(\Delta^i_k)$. First,
after examining a wide range of experimental spectra, we proposed
that the excited states are coupled to quasiparticles $\Delta^i_k
\leftrightarrow E^i_k$, which modifies the final gap equation,
eq.~(\ref{Egapfun}). Once in the superconducting state, these QP
excitations become well defined, revealing the characteristic
peak-dip structures in the spectral function. The non-retarded
energy-gap function described herein is thus the direct consequence
of this QP-mediated interaction.

Secondly, the pair-pair interaction also depends directly on the
condensate\,: $\beta(T) = \beta_0\,N_{oc}(T)$, which we identify as
the order parameter. Not only does it vanish at $T_c$, but it is
directly linked to long-range order. This interaction varies
spatially in the inhomogeneous superconductor, but also as a
function of concentration $p$. In the latter case, the interaction
is larger with increasing $p$ which leads to a {\it decrease} in the
self-consistent gap $\Delta_p$ and consequently to the lowering of
$T_c$ as well. Paradoxically, the very potential that is responsible
for superconducting order eventually, in the high-density limit,
provokes its demise.


\end{document}